\documentclass[twocolumn]{aastex631}

\shorttitle{Tidal truncation of circumplanetary disks}
\shortauthors{R. G. Martin et al.}
\begin{document}

\title{Tidal truncation of circumplanetary disks fails above a critical disk aspect ratio}

\author[0000-0003-2401-7168]{Rebecca G. Martin}
\affil{Nevada Center for Astrophysics, University of Nevada, Las Vegas,
4505 South Maryland Parkway, Las Vegas, NV 89154, USA}
\affil{Department of Physics and Astronomy, University of Nevada, Las Vegas,
4505 South Maryland Parkway, Las Vegas, NV 89154, USA}
\author[0000-0001-5032-1396]{Philip J. Armitage}
\affiliation{Center for Computational Astrophysics, Flatiron Institute, 162 Fifth Avenue, New York, NY 10010, USA}
\affiliation{Department of Physics and Astronomy, Stony Brook University, Stony Brook, NY 11794, USA}
\author[0000-0002-4636-7348]{Stephen H. Lubow}
\affiliation{Space Telescope Science Institute, 3700 San Martin Drive, Baltimore, MD 21218, USA}
\author[0000-0002-4716-4235]{Daniel J. Price}
\affiliation{School of Physics and Astronomy, Monash University, Vic. 3800, Australia}

%% Note that the \and command from previous versions of AASTeX is now
%% depreciated in this version as it is no longer necessary. AASTeX 
%% automatically takes care of all commas and "and"s between authors names.

%% AASTeX 6.2 has the new \collaboration and \nocollaboration commands to
%% provide the collaboration status of a group of authors. These commands 
%% can be used either before or after the list of corresponding authors. The
%% argument for \collaboration is the collaboration identifier. Authors are
%% encouraged to surround collaboration identifiers with ()s. The 
%% \nocollaboration command takes no argument and exists to indicate that
%% the nearby authors are not part of surrounding collaborations.

%% Mark off the abstract in the ``abstract'' environment. 
\begin{abstract}
We use numerical simulations of circumplanetary disks to determine the boundary between disks that are radially truncated by the tidal potential, and those where gas escapes the Hill sphere. We consider a model problem, in which a coplanar circumplanetary disk is resupplied with gas at an injection radius  smaller than the Hill radius. We evolve the disk using the {\sc Phantom} Smoothed Particle Hydrodynamics code until a steady-state is reached. We find that the most significant dependence of the truncation boundary is on the disk aspect ratio $H/R$. Circumplanetary disks are efficiently truncated for $H/R \lesssim 0.2$. For $H/R \simeq 0.3$, up to about half of the injected mass, depending on the injection radius, flows outwards through the decretion disk and escapes.
As expected from analytic arguments, the conditions ($H/R$ and Shakura-Sunyaev $\alpha$) required for tidal truncation are independent of planet mass.
A simulation with larger $\alpha=0.1$ shows stronger outflow than one with $\alpha=0.01$, but the dependence on transport efficiency is less important than variations of $H/R$. 
Our results suggest two distinct classes of circumplanetary disks: tidally truncated thin disks with dust-poor outer regions, and thicker actively decreting disks with enhanced dust-to-gas ratios. Applying our results to the PDS~70c system, we predict a largely truncated circumplanetary disk, but it is possible that enough mass escapes to support an outward flow of dust that could explain the observed disk size. 
\end{abstract}

%% Keywords should appear after the \end{abstract} command. 
%% See the online documentation for the full list of available subject
%% keywords and the rules for their use.

%% From the front matter, we move on to the body of the paper.
%% Sections are demarcated by \section and \subsection, respectively.
%% Observe the use of the LaTeX \label
%% command after the \subsection to give a symbolic KEY to the
%% subsection for cross-referencing in a \ref command.
%% You can use LaTeX's \ref and \label commands to keep track of
%% cross-references to sections, equations, tables, and figures.
%% That way, if you change the order of any elements, LaTeX will
%% automatically renumber them.
%%
%% We recommend that authors also use the natbib \citep
%% and \citet commands to identify citations.  The citations are
%% tied to the reference list via symbolic KEYs. The KEY corresponds
%% to the KEY in the \bibitem in the reference list below. 

\section{Introduction}
The outer extent of accretion disks in binary systems is typically limited by the effects of tidal forces from the companion. For geometrically thin, turbulent, disks, the truncation radius is set by a balance between tidal and viscous forces \citep{Papaloizou1977}, and normally lies close to the location where free particle orbits first intersect \citep{Paczynski1977}. Observations of disks in dwarf novae \citep{Smak84} are broadly consistent with these considerations, which have been tested using simulations and extended to the case of eccentric and inclined disks \citep{Artymowicz1994,Larwood96,Lubow15,Miranda2015}.

\cite{Martin2011} determined the tidal
truncation radius of a coplanar circumplanetary disk to be about $0.4 R_{\rm H}$, where the Hill radius for a planet of mass $M_{\rm p}$ orbiting at radius $a_{\rm p}$ from a star of mass $M_{\rm s}$ is $R_{\rm H}=(M_{\rm p}/(3(M_{\rm s}+M_{\rm p})))^{1/3}a_{\rm p}$, based on where viscous and tidal torques are in balance. Beyond this radius, free particle orbits intersect or become dynamically unstable. \cite{Quillen1998} estimated the truncation radius to be $R_{\rm H}/3$, based on the cicularization radius of an inflowing gas stream. This estimate neglects the outward viscous expansion of the disk required for steady accretion onto the planet. Both estimates are coincidentally numerically similar. 

The predicted structure of circumplanetary disks differs qualitatively depending on whether or not efficient tidal truncation occurs. Three dimensional simulations of the feeding of circumplanetary disks, from larger-scale protoplanetary disks, show that gas is replenished from out-of-plane flows \citep{DAngelo2003,Ayliffe2009,Tanigawaetal2012,Gressel13,Szulagyi14,Schulik20} that intersect the circumplanetary disk at a small fraction of the Hill radius. 
If such a disk is tidally truncated, the appropriate approximate analytic description matches a steady-state accretion disk interior to the mass injection radius, with a static outer disk that has the structure (density distribution) of a decretion disk \citep{Pringle1991,Nixon2021}, but in which the radial velocity is everywhere zero \citep{Martin2011}.  If truncation fails, the appropriate description of the outer disk is also a decretion disk but with an outwardly directed mean gas flow \citep{Ward10}. Gas outflow could counteract the otherwise inward drift of aerodynamically coupled solid particles \citep{Whipple72,Weidenschilling77aero}, allowing an actively decreting disk to accumulate a much larger mass of solids than one that is tidally truncated \citep{Batygin2020}.

In addition to the implications for satellite formation, which were the focus of \citet{Batygin2020} and multiple prior works \citep[examples include][]{Canup02,Mosqueira03}, differences in the gas and dust content of truncated and non-truncated disks are potentially observable.
The circumplanetary disk around PDS~70c \citep{Isella19,Benisty21} has an inferred dust mass of $7 \times 10^{-3} M_\oplus - 3 \times 10^{-2} M_\oplus$ \citep[depending on the assumed grain size,][]{Benisty21}. For the candidate cicumplanetary disk at large radius (200~au) in the AS~209 system, molecular line emission suggests a gas mass $\gtrsim 30 M_\oplus$, while limits on continuum emission imply a dust-to-gas ratio at least an order of magnitude below the fiducial value of $10^{-2}$ \citep{Bae22}.

Truncation of circumplanetary disks remains in doubt because their predicted aspect ratios, $H/R$, span a substantial range and extend to larger values than disks in most other mass transfer binaries. Consider a circumplanetary disk heated by stellar irradiation to the same temperature as the surrounding protoplanetary disk. At 0.4~Hill radii, where truncation would nominally occur, the aspect ratio of a circumplanetary disk around a Jupiter mass planet is about five times greater than that of the protoplanetary disk, giving $(H/R)\simeq~0.2-0.3$. Similar values are predicted for actively accreting disks \citep{Martin2011}.  High viscosity and non-negligible thermal pressure in such thick disks mean that efficient tidal truncation may not always occur.

In this work our goal is to determine the boundary between truncated and non-truncated circumplanetary disk solutions. We consider an idealized model of an isolated planet on a circular orbit, with a non-inclined circumplanetary disk fed steadily with gas at a specified injection radius. We evolve the hydrodynamic system to steady state, and assess the resulting efficiency of tidal truncation as a function of $H/R$. We further vary the viscosity, planet mass and injection radius in order to understand why truncation breaks down in thick disks.

Section \ref{sec:wotrunc} describes a 1D analytic model for circumplanetary disk flow that ignores the effects of the tidal field. Section \ref{sec:CPDSim} describes the results of 3D simulations of the circumplanetary disk flow.
Section \ref{conc} contains the discussion and conclusions.

\begin{table*}
\begin{center}
\begin{tabular}{l c c c c c c c c c c c c c}
\hline
 Name & $H/R$ & $\alpha_{\rm AV}$ & $M_{\rm p}/M_{\rm s}$ & $R_{\rm inj}/R_{\rm H}$ & $v_{\rm inj}/v_{\rm K}$&$M_{\rm part}/M_{\rm s}$  & $M_{\rm steady}/M_{\rm s}$ & $N_{\rm steady}$ & $\left< h\right>/H$ & $\alpha$ & $\dot M_{\rm p}/\dot M_{\rm inj}$ \\
 \hline
\hline
%run41again2
sim1a & 0.3 & 0.25 & 0.001 & 0.0064 & 1.00 & $1\times 10^{-12}$  & $2.56\times 10^{-8}$ & 26,000 & 0.43 &0.011 &  0.93 
\\
%sim41bagain 
sim1b & 0.3 & 0.41 & 0.001 & 0.0064 & 1.00& $5\times 10^{-13}$  & $2.76\times 10^{-8}$ & 55,000 &0.33 &0.014 &  0.93  \\
%sim41ca 
sim1c & 0.3 & 0.41  & 0.001 & 0.0064& 1.00 & $2\times 10^{-13}$  & $3.24\times 10^{-8}$ & 162,000 & 0.24 & 0.010 &  0.93  \\
%sim41d
sim1d & 0.3 & 0.41 & 0.001 & 0.0064& 1.00 & $1\times 10^{-13}$  & $3.71\times 10^{-8}$ & 371,000 & 0.18 & 0.008 &  0.93  \\
\hline
%run45g
sim2 & {\bf 0.1} & 0.24  & 0.001 & 0.0064& 1.00 & $5\times 10^{-12}$  & $2.22\times 10^{-7}$ & 44,000 & 0.62 & 0.015 & 1.00\\
%run44f
sim3  & {\bf 0.2} & 0.25 & 0.001 & 0.0064& 1.00 & $1\times 10^{-12}$  & $6.45\times 10^{-8}$ & 65,000 & 0.40 & 0.010 &  0.97 \\ 
%sim42c
sim4 & {\bf 0.4} & 0.63 & 0.001 & 0.0064& 1.00 & $2\times 10^{-13}$  & $2.89\times 10^{-8}$ & 144,000 & 0.20 & 0.013 &  0.85  \\
\hline
%sim47c
sim5 & 0.3 & 0.50 & {\bf 0.01} & 0.0064 & 1.00& $2\times 10^{-13}$ & $3.89\times 10^{-8}$ & 195,000 & 0.22 &0.011 & 0.90 \\
\hline
%run14h
sim6 & {\bf 0.1} & 0.24 & 0.001 & {\bf 0.14} & 1.00& $2\times 10^{-11}$  & $9.98\times 10^{-7}$ & 50,000 &0.55 & 0.013 &  $1.00$\\
% sim15d
sim7 & {\bf 0.2} & 0.50 & 0.001 & {\bf 0.14}& 1.00 &$ 1\times 10^{-12}$  & $2.82 \times 10^{-7}$& 281,000 & 0.21 & 0.011 & $0.87$\\
%sim16b
sim8 & 0.3 & 0.41 & 0.001 & {\bf 0.14}& 1.00 &$1\times 10^{-12}$  & $9.83\times 10^{-8}$  & 98,000 & 0.23 & 0.009 &  $0.53$\\
sim8b & 0.3 & 0.41 & 0.001 & \bf 0.14 & \bf 0.94  & $1\times 10^{-12}$  & $9.42\times 10^{-8}$  & 94,000 & 0.24 & 0.010 &  $0.62$\\
\hline
%run21f
sim9 & {\bf 0.1} & 2.00  & 0.001 & {\bf 0.14} & 1.00& $5\times 10^{-12}$  & $5.23\times 10^{-7}$ & 105,000 &0.44 & 0.088 &  1.00 \\
%sim23
sim10 & {\bf 0.2} & 3.10 & 0.001 & {\bf 0.14} & 1.00&$1\times 10^{-12}$  & $1.05\times 10^{-7}$ & 105,000 & 0.30 & 0.092 & 0.80  \\
\hline
%run50f
sim11 & {\bf 0.1} & 0.15 & 0.001 & {\bf 0.33} & 1.00& $5\times 10^{-11}$ & $1.19\times 10^{-6}$ & 24,000 & 0.66 & 0.010 & 1.00 \\
% run51
sim12 & {\bf 0.2} & 0.50 & 0.001 & {\bf 0.33}& 1.00 &$ 1\times 10^{-12}$ & $2.64\times 10^{-7}$ &  264,000& 0.22 & 0.011 & 0.74 \\
\hline
\end{tabular}
\end{center}
 \caption{Simulation parameters. Columns (left to right, respectively) are the simulation name; disk aspect ratio; SPH artificial viscosity parameter, $\alpha_{\rm AV}$; mass of the planet; injection radius; azimuthal injection velocity; mass of the SPH particles; mass of the steady disk; number of SPH particles; smoothing length in the steady disk; calculated value for the viscosity parameter in the steady disk; and finally the accretion rate on to the planet as a fraction of the injection accretion rate.
}
\label{table2}
\end{table*}

\section{Circumplanetary disk model without tidal truncation}
\label{sec:wotrunc}

Before considering the effects of tidal forces on  circumplanetary disks, we first describe an analytic model that omits the tidal field.
In this model we choose a simple prescription for the injection of material into the circumplanary disk. Material is injected into an axisymmetric viscous accretion disk at a single radius, $R_{\rm inj}$, with the Keplerian velocity, as in our simulations described in Section \ref{sec:CPDSim}. The process of accretion in to the circumplanetary disk is complicated by uncertainties involving the degree of magnetic and viscous angular momentum transfer in meridional flows within the Hill sphere \citep[e.g.][]{Szulagyi14}. Furthermore, in our model, material freely flows onto the planet, however, we note that the process of accretion onto the planet itself may be magnetically regulated \citep[e.g.][]{Adams2012}.  Interior to the injection radius mass is accreted towards the planet, while mass is decreted exterior to the injection radius.
Our analytic model is similar to that of \cite{Canup02}, except that they considered mass injection over a range in radii.
The flow model in \cite{Batygin2020} was based on a pure outflow (decretion) disk model. 
The inner and outer boundaries are taken to be at a radius near the planet and the radius of the Hill sphere, respectively.
As in \cite{Canup02}, we  apply a zero stress boundary condition at both the inner and outer boundaries. In doing so there
needs to be an extra degree of freedom for the equations to be properly constrained. That degree of freedom is the ratio of the mass flow rates at the inner and outer boundaries. 

We ignore the tidal field and apply the 1D equations (61)--(63) of \citet{Martin2011} with mass flux $F$ and injected mass flux $F_{\rm inj}>0$. We have
\begin{equation} 
\frac{d F}{dR} = F_{\rm inj} \delta(R-R_{\rm inj})
\label{dmdR}
\end{equation}
and
\begin{equation}
\frac{d (F R^2 \Omega)}{dR} = \frac{d T_{\rm v}}{dR} +  F_{\rm inj} R_{\rm inj}^2 \Omega_{\rm \rm inj} \delta(R-R_{\rm inj}),
\label{djdR}
\end{equation}
where $T_{\rm v}$ is the usual viscous stress
\begin{equation}
T_{\rm v} = - 3 \pi \nu \Sigma R^2 \Omega.
\end{equation}

We integrate Equation (\ref{dmdR}) to obtain
\begin{equation}
F_{\rm o} - F_{\rm i} = F_{\rm inj},
\end{equation}
where  $F_{\rm i} <0$  ($F_{\rm o} >0)$ is the mass flux interior (exterior) to the injection radius.
We integrate Equation  (\ref{djdR}) to obtain 
\begin{eqnarray}
T_{\rm v} &= F_{\rm i} R^2 \Omega +C_1, &~~~  R < R_{\rm inj}, \\
T_{\rm v} &= F_{\rm o}  R^2 \Omega - F_{\rm inj} R_{\rm inj}^2 \Omega_{\rm \rm inj} +C_2, &~~~  R> R_{\rm inj}.
\end{eqnarray}
Applying these equations at $R_{\rm inj}$ and requiring continuity of $T_{\rm v}$,  we have that
\begin{equation}
C_1=C_2.
\end{equation}
We also apply boundary conditions $T_{\rm v}(R_{\rm i})=0$ and $T_{\rm v}(R_{\rm o})=0$. We adopt the notation that $F_{\rm i}= -\dot{M}_{\rm p}$, where $\dot M_{\rm p}$ is the accretion rate on to the planet, $F_{\rm inj}= \dot{M}_{\rm inj}$, $F_{\rm o}=\dot{M}_{\rm o}$ so that $\dot{M}_{\rm inj}=\dot{M}_{\rm p}+\dot{M}_{\rm o}$
and obtain
\begin{equation}
\frac{\dot{M}_{\rm p}}{\dot{M}_{\rm inj}}= \frac{ \sqrt{R_{\rm o}} - \sqrt{R_{\rm inj}}}{\sqrt{R_{\rm o}} - \sqrt{R_{\rm i}}}
\label{Mdin}
\end{equation}
and
\begin{equation}
\frac{\dot{M}_{\rm o}}{\dot{M}_{\rm inj}}= \frac{ \sqrt{R_{\rm inj}} - \sqrt{R_{\rm i}}}{\sqrt{R_{\rm o}} - \sqrt{R_{\rm i}}}.
\label{Mdout}
\end{equation}
Note that for $R_{\rm i} \ll R_{\rm inj} \ll R_{\rm o}$, we have that 
\begin{equation}
\frac{\dot{M}_{\rm o}}{\dot{M}_{\rm inj}}= \sqrt{\frac{R_{\rm inj}}{R_{\rm o}}}
\end{equation}
that is similar to the first line of equation (A10) of \citet{Canup02}.
Equations~(\ref{Mdin}) and~(\ref{Mdout}) show that  the mass outflow rate increases with increasing injection radius,
while the mass inflow rate towards the planet increases with decreasing  injection radius.

The analytic model provides a rough indication of the effect of changing the injection radius on the accretion rate on to the planet. There are some differences between this analytic model and the numerical simulations presented in the next section. Unlike the numerical simulations, the analytic model ignores the tidal field and gas pressure and consequently assumes that the gas orbits at Keplerian velocity. 

\section{Circumplanetary disk simulations}
\label{sec:CPDSim}

In this Section we determine the long-term steady state structure for a circumplanetary disk including the tidal effects from the star. We use the smoothed particle hydrodynamics \citep[SPH, e.g.][]{Monaghan1992,Price2012a}
%\djp{added M92} 
code {\sc phantom} \citep{Price2010,Lodato2010,Price2018}. Tidal truncation of accretion disks has been extensively studied using {\sc phantom} \citep[e.g.][]{Franchini2019inner,Heath2020,Hirsh2020}.

\subsection{Parameters and resolution study}
\label{sec:param}

We begin by considering a planet with mass $M_{\rm p}=0.001
\,M_{\rm s}$  orbiting a star of mass $M_{\rm s}$ in a circular orbit with semi-major axis $a_{
\rm p}$. The planet and the star are modelled as sink particles that accrete gas particles that fall within their radius \citep[e.g.][]{Bateetal1995}. The size of the planet is $R_{\rm acc}=0.0013\,R_{\rm H}$. In some cases we instead use  $R_{\rm acc}=0.028\,R_{\rm H}$. We inject particles at a radius of $R_{\rm inj}$ from the planet with Keplerian velocity around the planet, where the Hill radius is $R_{\rm H}=a_{\rm p}(M_{\rm p}/(3(M_{\rm s}+M_{\rm p})))^{1/3}$.  We consider three injection radii in our simulations. The first is very close to the planet, $R_{\rm inj}=0.0064\,R_{\rm H}$, and is motivated by 3D simulations of circumplanetary disk accretion \citep{Tanigawaetal2012,Morbidelli2014,Szulagyi2016}. The second value we take is much larger, $R_{\rm inj}=0.14\, R_{\rm H}$. The third is $R_{\rm inj}=0.33\, R_{\rm H}$ and is motivated by the angular momentum of gas flowing in the orbital plane from the protoplanetary disk to the circumplanetary disk, as discussed in the Introduction \citep[e.g.][]{Quillen1998,Ayliffe2009b}.  
The particle orbits of the injected particles are initially coplanar to the orbit of the planet around the star. Similar methods have previously  been used for binary star simulations with nearly equal mass ratio binaries \citep[e.g.][]{Okazaki2002,Suffak2022}. Particles are injected at a rate of $\dot M_{\rm inj}=1.2\times 10^{-7} M_{\rm s}/P_{\rm orb}$, where $P_{\rm orb}$ is the orbital period of the planet.  The parameters that we have chosen represent an injection radius that is equal to $5\,R_{\rm J}$, an injection accretion rate of $M_{\rm inj}=1\times 10^{-8}\,\rm M_\odot \, yr^{-1}$, and a sink size of 1 Jupiter radius for a Jupiter mass planet orbiting at $5.2\,\rm au$. \cite{Batygin2020} adopted the same injection radius but a lower accretion rate resulting in $H/R=0.1$. We implement the disk viscosity by adapting the SPH artificial viscosity. The values for $\alpha_{\rm AV}$ in each simulation are shown in Table~\ref{table2} and we take $\beta_{\rm AV}=2$. We calculate the \cite{SS1973} $\alpha$ in the steady disk with equation 38 in \cite{Lodato2010}. Although we can pick $\alpha_{\rm AV}$, we cannot easily control $\alpha$ because it depends on the smoothing length $\left<h\right>$ and in turn the disk density in the steady state disk.

 We evolve each simulation until the number of particles in $R<R_{\rm H}$ is constant in time. Table~\ref{table2} lists the approximate number of particles within the Hill sphere at this time, $N_{\rm steady}$. 
  Models sim1a - sim4 use $R_{\rm acc} = 0.0013 R_{\rm H}$ and other models use $R_{\rm acc} = 0.028 R_{\rm H}$.
The final column of Table~\ref{table2} shows the accretion rate onto the planet scaled to the mass injection rate, $\dot M_{\rm p}/\dot M_{\rm inj}$.  
We refer to this ratio as the {\it tidal truncation efficiency}.

The disk aspect ratio is constant with radius in all simulations. We first consider four simulations with $H/R=0.3$ with different SPH particle masses, sim1a, sim1b, sim1c and sim1d (see Table~\ref{table2}). The mass of the disk increases with the increasing resolution. However, the accretion rate  on to the planet does not change significantly in the steady state disk.   

\subsection{Effect of the disk aspect ratio}

\begin{figure}
\begin{center}
\includegraphics[width=\columnwidth]{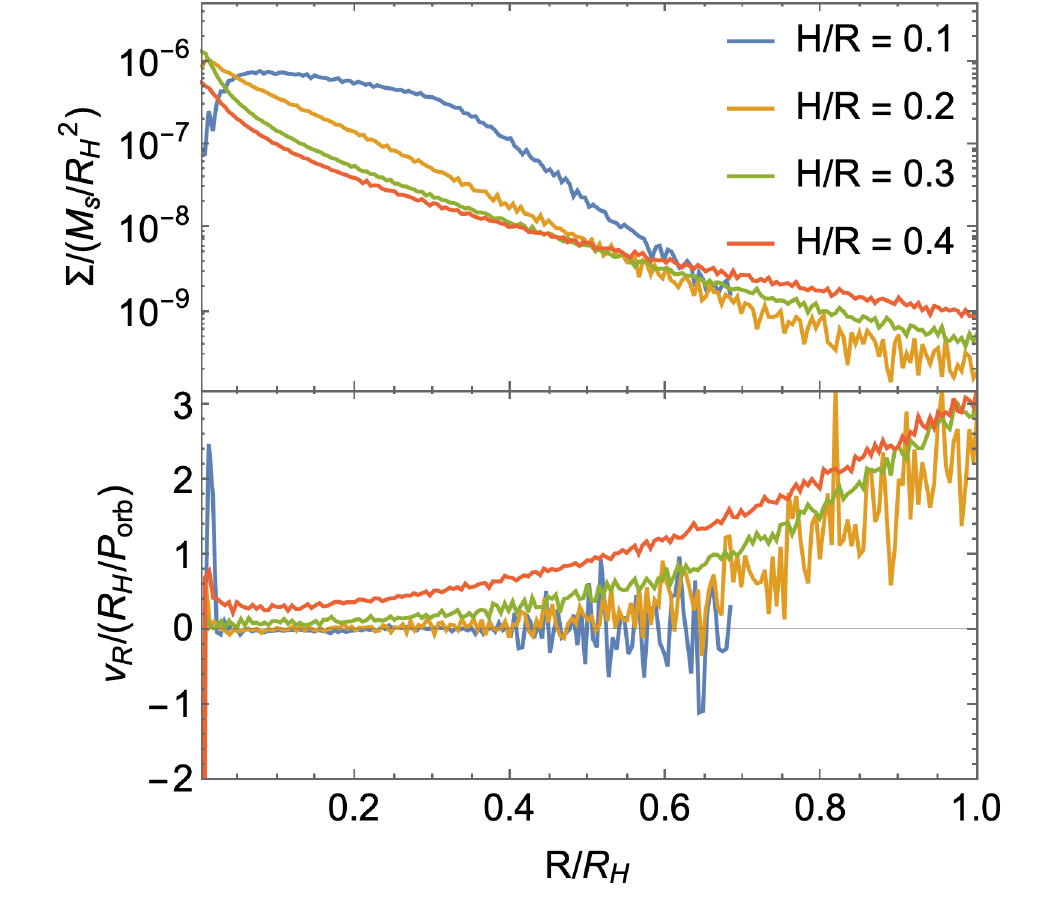} 
	\end{center}
    \caption{The steady state surface density distribution  (upper) and radial velocity (lower) of the disk for disk aspect ratio $H/R=0.1$ (sim2), 0.2 (sim3), 0.3 (sim1c) and 0.4 (sim4). 
    %The dashed vertical line shows the injection radius. {\bf Maybe just say what the injection radius is and remove the vertical lines. They are hard to see on the plot. } 
    }
    \label{mdot}
\end{figure}

\begin{figure*}
\begin{center}
\includegraphics[width=0.8\textwidth]{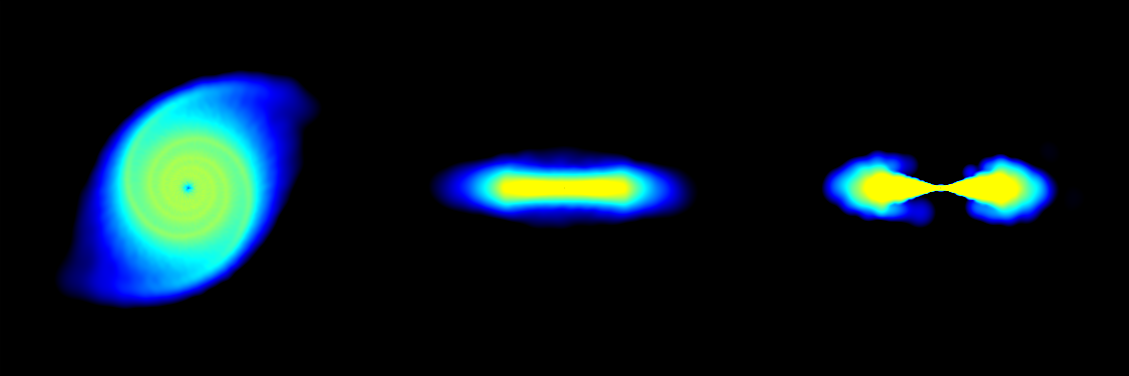}
\includegraphics[width=0.8\textwidth]{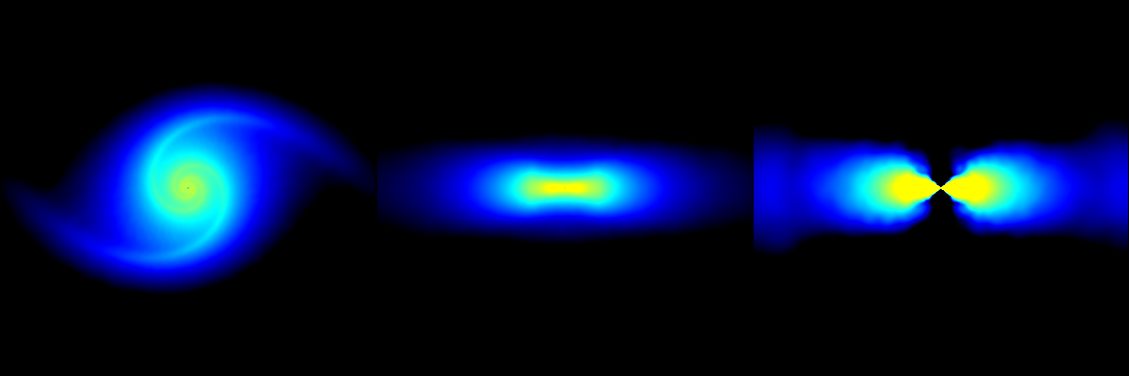}
\includegraphics[width=0.8\textwidth]{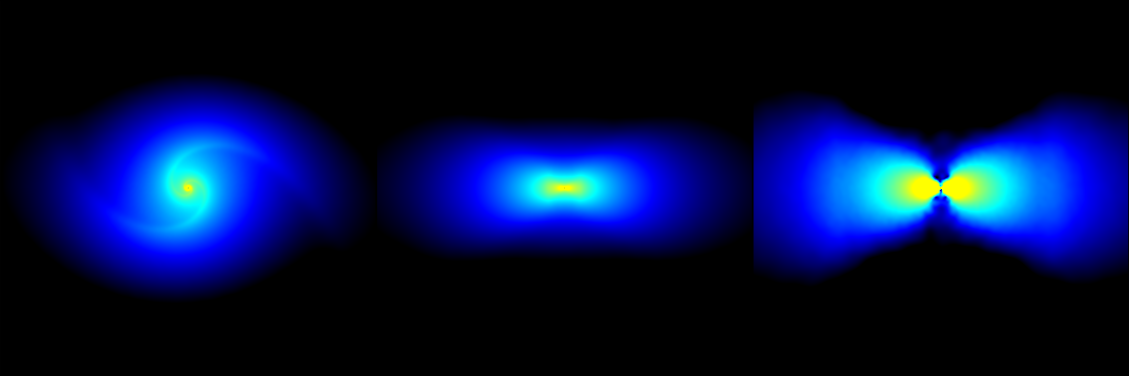} 
\includegraphics[width=0.8\textwidth]{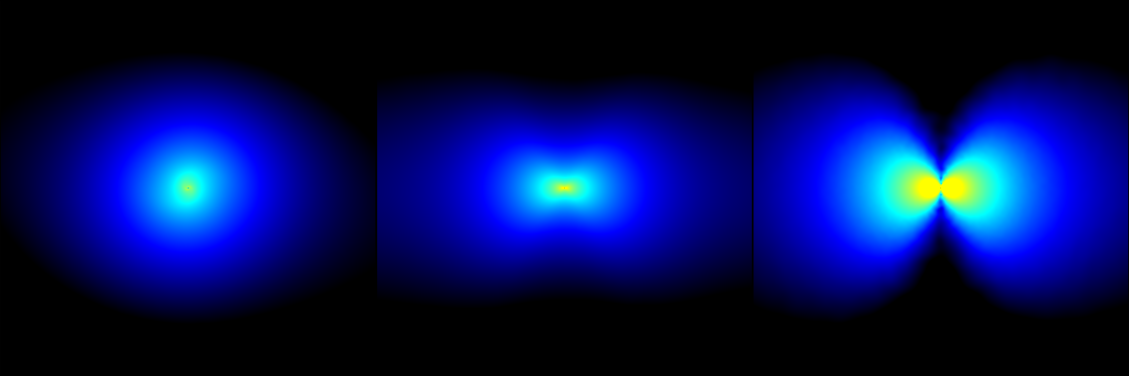} 
	\end{center}
    \caption{Visualizations of the final density distribution in selected simulations. From left to right, the panels show the planet orbital plane (the $x-y$ plane left), the $x-z$ plane (middle) and a cross section in the $x-z$ plane (right). The  planet is at the center of each panel. The size of each plot is twice the Hill radius. The star is not shown but is along the positive $x$-axis, to the right.  From top to bottom the disk aspect ratio is $H/R=0.1$ (top, sim2), 0.2 (second row, sim3), 0.3 (third row, sim1c), and 0.4 (bottom row, sim4), with $\alpha\approx 0.01$. 
}  
    \label{splash}
\end{figure*}

We now consider a range of disk aspect ratios from 0.1 to 0.4 as shown in Table~\ref{table2}. Fig.~\ref{mdot} shows the surface density and radial velocity profiles while
Fig.~\ref{splash} shows the density distributions at the end of the simulations for different values of $H/R$ with fixed $\alpha\approx 0.01$ and injection radius $R_{\rm inj}=0.0064 R_{\rm H}$. 
Material that is injected around the planet spreads outwards through the effects of viscosity, at least initially. 

For disk aspect ratio $H/R = 0.1$, the tidal truncation efficiency is 100\%. 
The tidal truncation is sufficiently strong to prevent material from escaping from the Hill sphere. The disk is an accretion disk for $R<R_{\rm inj}$ and has zero radial velocity for $R>R_{\rm inj}$. The tidal effects from the star remove angular momentum from the material in the outer parts of the disk. The tidal torque on the disk is associated with
 spiral arms \citep{Martin2011, Zhu2016}, as seen in Fig.~\ref{splash}.
Apart from the innermost region, the density declines slowly in radius for smaller $R$ and declines rapidly beyond the expected truncation radius at $R=0.4 R_{\rm H}$. 

As $H/R$ increases, a larger fraction of material can escape the tidal potential. For $H/R \ge 0.2$, the density profiles decline in radius without a major change in slope at the truncation radius.  For $H/R \gtrsim 0.3$, there is still a disk structure, but the tidal torque is not sufficiently strong to prevent significant material from escaping from the Hill sphere. The tidal truncation efficiency is about 85\% for $H/R=0.4$. This means that there is a significant density of material extending to the Hill radius for large values of $H/R$.  For $H/R = 0.4$, the disk structure is quite different. There are no longer spiral arms in the disk 
and the material flows outwards in a thick flow (see the lower  panel of Fig.~\ref{splash}). 
This issue is related to the reduction of the tidal torques with increasing disk aspect ratio as discussed in Section~\ref{2.6}. But a disk with $H/R=0.4$ is more torus than disk!

For a fixed injection rate, the circumplanetary disk density is found to be higher in the $H/R=0.1$ case (that has truncation efficiency of 1)
 than for thicker disks  up to a radius of around $0.6-0.7 R_{\rm Hill}$. This suggests that despite a lower truncation efficiency and the material spreading towards larger radii, thick circumplanetary disks might appear fainter and smaller due to their lower masses and densities.

\subsection{Effect of the planet mass}

\begin{figure}
\begin{center}
\includegraphics[width=\columnwidth]{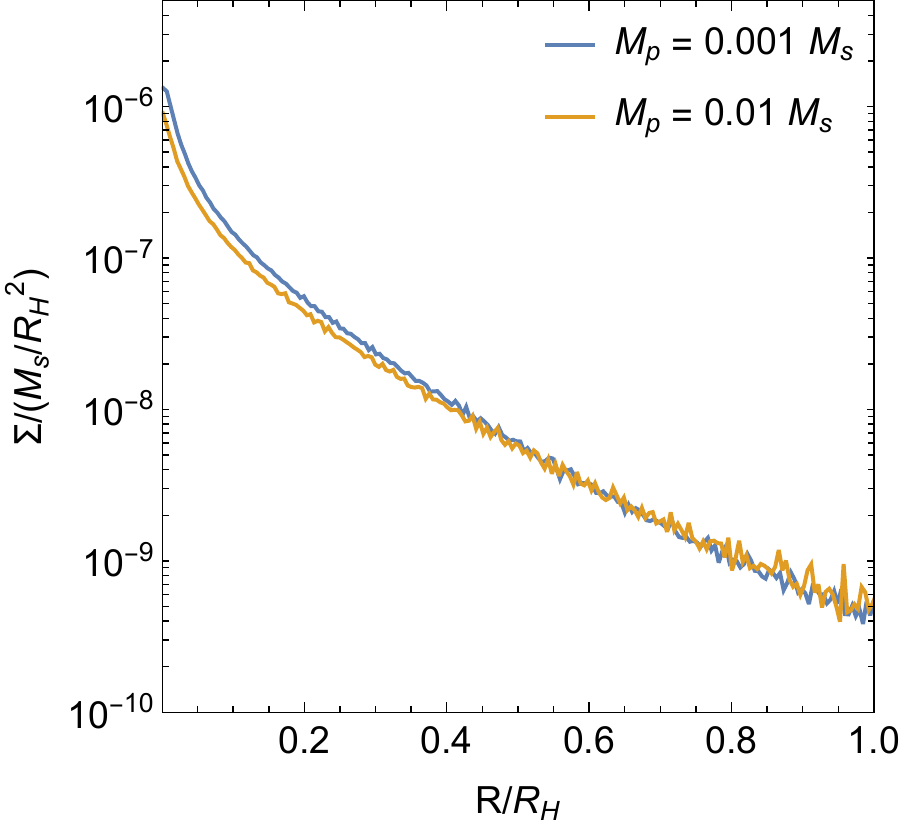} 
	\end{center}
    \caption{ The effect of the planet mass on the surface density structure for models with $H/R=0.3$ and $\alpha \approx 0.01$. The planet mass is $0.001\,M$ (sim1c) for the blue line and $0.01\, M$ (sim5) for the yellow line. The radius is scaled to the respective Hill radius in each case. 
    %The vertical dashed line shows the particle injection radius.
    }  
    \label{fig:planetmass}
\end{figure}

We now consider a simulation with a larger planet mass of $0.01\, M_{\rm s}$ (sim5) with $H/R=0.3$ and $\alpha=0.01$ and compare this to sim1c with the same disk parameters. 
The equations of motion for a circumplanetary test particle can be scaled 
to be independent of planet mass, provided that the planet mass is much smaller than the stellar mass in the Hill approximation.
 The equations
 of motion for a viscous circumplanetary disk can be similarly scaled to be independent of planet mass, provided that the fluid properties $H/R$ and $\alpha$
are fixed \citep{Martin2011}. In the current case with mass injection, the injection radius is also scaled to be fixed relative to the Hill sphere radius. Under these conditions, the scaled gas surface density distribution $\Sigma(R/R_{\rm H})/R_{\rm H}^2$ should be independent of planet mass.
Fig.~\ref{fig:planetmass} shows the
surface density of the steady state disks both scaled to their respective Hill radii. Changing the planet mass does not affect the surface density profile scaled to the Hill radius, as expected.
In addition, the tidal truncation efficiency should be independent of planet mass. We recover this predicted result.

\subsection{Effect of the injection radius }
\label{sec:injrad}

We now compare the results of simulations with  injection radii of $R_{\rm inj}=0.0064$, $0.14$ and $0.33\,R_{\rm H}$. With these injection radii, models sim2, sim6, and sim11 in Table~\ref{table2} show that for a thin disk  with $H/R=0.1$,
the truncation is perfectly efficient, $\dot{M}_{\rm p}/\dot{M}_{\rm inj} = 1$, independent of the injection radius.
This high efficiency is due to the dominance of the tidal effects of the star over the effects of disk gas pressure and viscosity.
The truncation efficiency is lower  for a thicker disk  with $H/R=0.2$ at these respective injection radii  in models sim3, sim7, and sim12. 
The simulations show that the truncation efficiency decreases with increasing injection radius. This trend is in qualitative agreement
with the analytic model predictions of Equation (\ref{Mdin}).

Since the analytic model contains a number of simplifying assumptions, it is difficult to make a quantitative comparison between the analytic model that contains the effects of viscosity and the numerical simulations that additionally  include the effects of the tidal field,  gas pressure, and the resulting non-Keplerian rotation.
Furthermore, it is difficult to disentangle the effects  tidal field,  gas pressure, and the non-Keplerian rotation
in the simulations. 
Nonetheless, we consider the application of Equation (\ref{Mdin}) to our simulation results that overrun the tidal field.  We set $R_{\rm i}=R_{\rm acc}$ (as described in Section \ref{sec:param}) and $R_{\rm o} = R_{\rm H}$.

Under the conditions of  sim1a-sim4, $R_{\rm acc}=0.0013 R_{\rm H}$ and $R_{\rm inj}=0.0064 R_{\rm H}$, we have from Equation (\ref{Mdin}) that $\dot{M}_{\rm p}=0.95\dot{M}_{\rm inj}$. 
The simulations show that the truncation efficiency  is high in the case of a small injection radius,
 even at larger values of $H/R$ when the effects of the tidal field are less dominant. For such small injection radii, the flow onto the planet cannot be ignored when considering the outcome of  the injected mass flow.
For sim3, sim7, and sim12 with $H/R=0.2$, the truncation efficiency is $\dot{M}_{\rm p}/\dot{M}_{\rm inj} = 0.95, 0.75,$ and 0.53, respectively, compared with the simulated values of 0.97, 0.87, and 0.74.
It is not surprising that the corresponding simulated values are larger because of the effects of the tidal field.
Notice that the truncation efficiency change in going from a model with $R_{\rm inj}=0.14\,R_{\rm H}$ to $0.33\,R_{\rm H}$ at fixed $H/R=0.2$ (sim7 and sim12), is much smaller than the efficiency change due to increasing $H/R$ in the models with $0.14\,R_{\rm H}$  from $H/R=0.2$ to to  0.3 (sim7 and sim8).

To compare the analytic estimate to a simulation  without a tidal field, we ran some numerical simulations without the star. The mass of the disk inside of the Hill sphere increased in time and the disk did not reach a steady state. We also ran some simulations in which we removed particles outside of the Hill sphere. In this case, the inner parts of the disk became steady, but there was nonsteady and nonaxisymmetric behaviour in the outer parts of the disk, outside of the injection radius. We speculate that this may be related to the \cite{Papaloizou1984} instability. This instability is not present when the star is included. Therefore we are unable to directly compare the analytic model to these simulations.

\subsection{Effect of the $\alpha$ viscosity parameter}
 
We now consider the effect of changing the $\alpha$ viscosity parameter. To test that dependence, we use the larger injection radius which leads to more disk outflow.    
We compare simulations sim7 and sim10 that have the same value of $H/R =0.2$, but with different values of $\alpha\approx 0.01$ and 0.1, respectively. 
The simulation with the higher $\alpha$  has about twice the rate of gas outflow, $1- \dot{M}_{\rm p}/\dot{M}_{\rm inj}$,
as the simulation with the lower $\alpha$. This change is in qualitative
agreement with expectations,  
since the viscous torque is stronger in the high $\alpha$ case and better able to overcome the tidal torque.
Nonetheless, this result suggests the
sensitivity of the outflow rate to $\alpha$ is not very
high, given that $\alpha$ has been increased by a factor of ten.
A higher sensitivity occurs to variations in $H/R$ at fixed $\alpha$, as can be seen by comparing models sim7 and sim8.
This therefore suggests that the value of $\alpha$ is not as important in determining the tidal truncation efficiency as $H/R$.

\subsection{Effects at fixed kinematic viscosity}
 \label{2.6}
 
Viscous disk truncation occurs when there is a balance between viscous and tidal toques, as discussed in the Introduction.
The viscous torque depends on the kinematic turbulent viscosity. Its dependence  on $H/R$ and $\alpha$
enters through the product $\alpha~(H/R)^2$. For a fixed tidal torque, the tidal truncation efficiency should  then depend on $\alpha~(H/R)^2$.
We performed a simulation with a higher value of $\alpha\approx 0.1$ and  disk aspect ratio of $H/R=0.1$ (sim9). This simulation has nearly the same effective kinematic viscosity as sim8 that has $\alpha~\approx~0.01$ and $H/R=0.3$.  However,  we find (see Table~\ref{table2}) that for the smaller $H/R$ case the tidal truncation efficiency is much higher than in the larger $H/R$ case. This result may suggest that the tidal torque depends on $H/R$. 

Such a dependence is possible because the
resonance that gives rise to the spiral arms seen in Fig. \ref{splash} does not lie within the circumplanetary disk \citep{Martin2011}. 
\cite{Xu2018} showed that in such cases the tidal torque
on the disk can depend on the sound speed, but their implied dependence on the sound speed (increasing torque with increasing sound speed) is opposite to what we infer.
 At sufficiently high values of $H/R$ the radial width of the resonance becomes comparable to the circumplanetary disk radius. 
The disk might then respond globally to the tidal field in a similar manner as occurs for (low frequency) secular resonances \citep{Lubow2001, Goldreich2003, Lubow2022} and not locally launch propagating waves. Under such conditions, the tidal torque might become weaker than in the thin disk case.
We will investigate these effects further in a future study. 

\subsection{Effect of the injection velocity}

The azimuthal velocity of the material in the disk becomes more sub-Keplerian with increasing disk aspect ratio. In the steady state disk, the azimuthal velocity at $R=R_{\rm inj}$ is $0.996\,v_{\rm K}$ in sim6, $0.96\,v_{\rm K}$ in sim7, and $0.94\,v_{\rm K}$ in sim8 where $v_{\rm K}$ is the local Keplerian velocity. However, the material flowing into the disk does not know what the disk aspect ratio is in the circumplanetary disk and therefore keeping the injection velocity constant (at Keplerian velocity) for the different simulations seems like the physically correct thing to do. However,  in order to disentangle the effect of increasing $H/R$ with the injection of material that is rotating faster than the disk we consider one final simulation. In sim8b we inject material at $0.94\,v_{\rm K}$ but everything else is the same as sim8.  The tidal truncation efficiency  is slightly higher in this case (0.62 compared to 0.53 in sim8), as would be expected. However, the effect of $H/R$ dominates the effect of changing the injection velocity.  

\section{Discussion and conclusions}
\label{conc}
We used hydrodynamic simulations of continuously fed circumplanetary disks to determine the conditions under which such disks are tidally truncated. Angular momentum {\em removal} from the circumplanetary disk occurs in our simulations through spiral shocks, which are resolved hydrodynamic features of the flow.
Angular momentum transport {\em within} the disk occurs through a viscous $\alpha$ torque, which is intended to represent transport from the Magnetorotational Instability \citep{Balbus98} or other turbulent processes \citep{Armitage2011,Lesur22}.
We found that the boundary, separating truncated disks from those where gas escapes the Hill sphere, depends primarily on the disk aspect ratio $H/R$ and the gas injection radius, and to a lesser extent on the the viscosity parameter $\alpha$. 

Almost complete truncation occurs for $H/R < 0.2$. At  $H/R = 0.4$ the disk structure is quite different: there are no spiral arms and the tidal field does not significantly restrict the outer regions of the disk. The truncation efficiency, defined as the fraction of injected gas that is ultimately accreted by the planet, depends both on $H/R$ and on the location of the injection radius. When truncation is incomplete, the fraction of mass that escapes is limited by angular momentum conservation, and is larger (at fixed $H/R$) for larger values of the injection radius. For example, the truncation efficiency drops to 50\% at $H/R \approx 0.3$, but only for a larger injection radius $R_{\rm inj}=0.14 R_{\rm H}$.
Consistent with the analytic arguments of \citet{Martin2011}, the truncation efficiency is independent of the planet mass. The truncation efficiency is more sensitive to changes in $H/R$ than $\alpha$.

On which side of the truncation boundary do real circumplanetary disks fall? The gaseous component of the circumplanetary disk around PDS~70c is currently unconstrained, but the dust continuum data is interpreted in terms of a disk with a temperature of 26~K, orbiting a 2~$M_{\rm J}$ planet 34~au from the host star. At $R_{\rm H}/3$, assuming a mean molecular weight $\mu =2.2$, these parameters would imply $H/R \approx 0.235$. Our results suggest that such a disk would be largely tidally truncated, but that it is possible that a small fraction of the injected mass would escape in a decretion flow that could advect dust outward. This is consistent (within large uncertainties) both with the detection of a spatially resolved source in the dust continuum data, and with the inferred spatial extent of 0.2-0.4~$R_{\rm H}$. We can also consider general scaling arguments. For an actively accreting disk, heated by viscous dissipation, \citet{Martin2011} estimated the aspect ratio to be,
\begin{eqnarray}
    \frac{H}{R} \simeq 0.3 \left( \frac{M_{\rm s}}{M_\odot} \right)^{1/24}
        \left( \frac{M_{\rm p}}{M_{\rm J}} \right)^{-1/3}
        \left( \frac{a}{5 \ {\rm au}} \right)^{1/8} \nonumber \\
        \times \left( \frac{\dot{M_{\rm p}}}{10^{-8} \ M_\odot {\rm yr}^{-1}} \right)^{1/8}
        \left( \frac{R}{R_{\rm H} / 3} \right)^{1/8}.
\end{eqnarray}
Here, the baseline accretion rate of $\dot{M}_{\rm p} = 10^{-8} \ M_\odot {\rm yr}^{-1}$ would lead to the accretion of a Jupiter mass in $10^5$~yr. We can also consider a passive disk, heated by stellar irradiation to the same temperature that the protoplanetary disk would have at the corresponding distance from the star. In this limit, the aspect ratio is given by,
\begin{equation}
    \frac{H}{R} \simeq 0.24 \left( \frac{H_{\rm pp} / a}{0.05} \right) 
        \left( \frac{M_{\rm p} / M_{\rm s}}{10^{-3}} \right)^{-1/3} 
        \left( \frac{R}{R_{\rm H} / 3} \right)^{1/2},      
\end{equation}
where $H_{\rm pp}/a$ is the aspect ratio that the {\em protoplanetary} disk would have at the orbital radius of the planet. While there is only a weak dependence on radius for the viscously heated disk, the passive disk may be quite flared and therefore have a smaller truncation efficiency. Both formulae have the same dependence on planet mass, in the sense that disks around more massive planets are (other factors being equal) more likely be tidally truncated. Although the estimates are uncertain---due for example to the unknown vertical profile of dissipation in the viscously heated case, and to the complex radiation field that would attend a circumplanetary disk in a gap in the irradiated case---it appears likely that real disks sit on both sides of the truncation boundary.

Comparing our results to the analytic model of \citet{Batygin2020}, which in part motivated this study, we confirm their most important qualitative physical conclusion: circumplanetary disks can support an active decretion flow that could allow the disk to accumulate solid material. Quantitatively, we find significant differences. First, for the specific parameters adopted by \citet{Batygin2020}, our simulations (in particular sim2) show that tidal truncation is efficient for lower values of $H/R\lesssim 0.2$. Second, even when truncation fails (indeed, even in the notional absence of any tidal field at all, see Section \ref{sec:injrad}), a significant fraction of the injected mass will be accreted on to the planet. The injection radius is a key parameter determining the rate of planetary accretion. For more realistic values of $H/R \gtrsim 0.3$, our results are in agreement with previous hydrodynamical simulations of the flow into a circumplanetary disk that show significant outflow \citep{Tanigawaetal2012,Morbidelli2014,Szulagyi2016}.

Our simulations treat the three-dimensional gas flows that feed circumplanetary disks crudely, and do not model dust or larger particles at all. Nonetheless, they suggest a physical picture in which there are (at least) two classes of circumplanetary disks, with substantially different properties. Sufficiently thin circumplanetary disks are fully truncated, and will support gas and dust inflow interior to the mass injection radius. Exterior to the mass injection radius the radial velocity in the gas will be zero, with only a tail of dust determined by a balance between inward radial drift and outward turbulent diffusion \citep{Clarke88}. Truncation fails for thicker disks with larger gas injection radii, whose outer regions will support an active decretion flow. 
Advection of particles within the decretion flow may allow higher solid-to-gas ratios to build up. Finally, we note that our simulations ignore magnetic fields. A magnetized circumplanetary disk \citep{Gressel13}, from which angular momentum is efficiently lost in a wind, mirroring recent protoplanetary disk models \citep{Bai2013,Pascucci22,Tabone22}, represents a third logical possibility. A wind-driven disk need not expand from the injection radius to the fiducial truncation radius at all, and would likely appear to be highly compact in both gas and dust tracers.

\begin{acknowledgements}
We thank anonymous referees for useful comments. Computer support was provided by UNLV’s National Supercomputing Center. RGM and SHL acknowledge support from NASA through grants 80NSSC21K0395 and 80NSSC19K0443. PJA and RGM acknowledge support from NASA TCAN award 80NSSC19K0639. 
We acknowledge the use of SPLASH \citep{Price2007} for the rendering of Fig.~\ref{splash}. The paper arose from our mutual interaction at the Kavli Institute for Theoretical Physics program on ``Bridging the Gap: Accretion and Orbital Evolution in Stellar and Black Hole Binaries", funded by the National Science Foundation under Grant No. NSF PHY-1748958. SHL thanks the Simons Foundation for support of a visit to the Flatiron Institute.
\end{acknowledgements}

%% The reference list follows the main body and any appendices.
%% Use LaTeX's thebibliography environment to mark up your reference list.
%% Note \begin{thebibliography} is followed by an empty set of
%% curly braces.  If you forget this, LaTeX will generate the error
%% "Perhaps a missing \item?".
%%
%% thebibliography produces citations in the text using \bibitem-\cite
%% cross-referencing. Each reference is preceded by a
%% \bibitem command that defines in curly braces the KEY that corresponds
%% to the KEY in the \cite commands (see the first section above).
%% Make sure that you provide a unique KEY for every \bibitem or else the
%% paper will not LaTeX. The square brackets should contain
%% the citation text that LaTeX will insert in
%% place of the \cite commands.

%% We have used macros to produce journal name abbreviations.
%% \aastex provides a number of these for the more frequently-cited journals.
%% See the Author Guide for a list of them.

%% Note that the style of the \bibitem labels (in []) is slightly
%% different from previous examples.  The natbib system solves a host
%% of citation expression problems, but it is necessary to clearly
%% delimit the year from the author name used in the citation.
%% See the natbib documentation for more details and options.

\bibliographystyle{aasjournal}
\bibliography{astroph} % if your bibtex file is called example.bib

\begin{thebibliography}{}
\expandafter\ifx\csname natexlab\endcsname\relax\def\natexlab#1{#1}\fi
\providecommand{\url}[1]{\href{#1}{#1}}

\bibitem[{{Adams} \& {Gregory}(2012)}]{Adams2012}
{Adams}, F.~C., \& {Gregory}, S.~G. 2012, \apj, 744, 55

\bibitem[{{Armitage}(2011)}]{Armitage2011}
{Armitage}, P.~J. 2011, \araa, 49, 195

\bibitem[{{Artymowicz} \& {Lubow}(1994)}]{Artymowicz1994}
{Artymowicz}, P., \& {Lubow}, S.~H. 1994, ApJ, 421, 651

\bibitem[{{Ayliffe} \& {Bate}(2009{\natexlab{a}})}]{Ayliffe2009}
{Ayliffe}, B.~A., \& {Bate}, M.~R. 2009{\natexlab{a}}, MNRAS, 393, 49

\bibitem[{{Ayliffe} \& {Bate}(2009{\natexlab{b}})}]{Ayliffe2009b}
---. 2009{\natexlab{b}}, MNRAS, 397, 657

\bibitem[{{Bae} {et~al.}(2022){Bae}, {Teague}, {Andrews}, {Benisty},
  {Facchini}, {Galloway-Sprietsma}, {Loomis}, {Aikawa}, {Alarc{\'o}n},
  {Bergin}, {Bergner}, {Booth}, {Cataldi}, {Cleeves}, {Czekala}, {Guzm{\'a}n},
  {Huang}, {Ilee}, {Kurtovic}, {Law}, {Gal}, {Liu}, {Long}, {M{\'e}nard},
  {{\"O}berg}, {P{\'e}rez}, {Qi}, {Schwarz}, {Sierra}, {Walsh}, {Wilner}, \&
  {Zhang}}]{Bae22}
{Bae}, J., {Teague}, R., {Andrews}, S.~M., {et~al.} 2022, \apjl, 934, L20

\bibitem[{{Bai} \& {Stone}(2013)}]{Bai2013}
{Bai}, X.-N., \& {Stone}, J.~M. 2013, \apj, 769, 76

\bibitem[{{Balbus} \& {Hawley}(1998)}]{Balbus98}
{Balbus}, S.~A., \& {Hawley}, J.~F. 1998, Reviews of Modern Physics, 70, 1

\bibitem[{{Bate} {et~al.}(1995){Bate}, {Bonnell}, \& {Price}}]{Bateetal1995}
{Bate}, M.~R., {Bonnell}, I.~A., \& {Price}, N.~M. 1995, \mnras, 277, 362

\bibitem[{{Batygin} \& {Morbidelli}(2020)}]{Batygin2020}
{Batygin}, K., \& {Morbidelli}, A. 2020, \apj, 894, 143

\bibitem[{{Benisty} {et~al.}(2021){Benisty}, {Bae}, {Facchini}, {Keppler},
  {Teague}, {Isella}, {Kurtovic}, {P{\'e}rez}, {Sierra}, {Andrews},
  {Carpenter}, {Czekala}, {Dominik}, {Henning}, {Menard}, {Pinilla}, \&
  {Zurlo}}]{Benisty21}
{Benisty}, M., {Bae}, J., {Facchini}, S., {et~al.} 2021, \apjl, 916, L2

\bibitem[{{Canup} \& {Ward}(2002)}]{Canup02}
{Canup}, R.~M., \& {Ward}, W.~R. 2002, \aj, 124, 3404

\bibitem[{{Clarke} \& {Pringle}(1988)}]{Clarke88}
{Clarke}, C.~J., \& {Pringle}, J.~E. 1988, \mnras, 235, 365

\bibitem[{{D'Angelo} {et~al.}(2003){D'Angelo}, {Henning}, \&
  {Kley}}]{DAngelo2003}
{D'Angelo}, G., {Henning}, T., \& {Kley}, W. 2003, \apj, 599, 548

\bibitem[{{Franchini} {et~al.}(2019){Franchini}, {Lubow}, \&
  {Martin}}]{Franchini2019inner}
{Franchini}, A., {Lubow}, S.~H., \& {Martin}, R.~G. 2019, \apjl, 880, L18

\bibitem[{{Goldreich} \& {Sari}(2003)}]{Goldreich2003}
{Goldreich}, P., \& {Sari}, R. 2003, ApJ, 585, 1024

\bibitem[{{Gressel} {et~al.}(2013){Gressel}, {Nelson}, {Turner}, \&
  {Ziegler}}]{Gressel13}
{Gressel}, O., {Nelson}, R.~P., {Turner}, N.~J., \& {Ziegler}, U. 2013, \apj,
  779, 59

\bibitem[{{Heath} \& {Nixon}(2020)}]{Heath2020}
{Heath}, R.~M., \& {Nixon}, C.~J. 2020, \aap, 641, A64

\bibitem[{{Hirsh} {et~al.}(2020){Hirsh}, {Price}, {Gonzalez},
  {Ubeira-Gabellini}, \& {Ragusa}}]{Hirsh2020}
{Hirsh}, K., {Price}, D.~J., {Gonzalez}, J.-F., {Ubeira-Gabellini}, M.~G., \&
  {Ragusa}, E. 2020, \mnras, 498, 2936

\bibitem[{{Isella} {et~al.}(2019){Isella}, {Benisty}, {Teague}, {Bae},
  {Keppler}, {Facchini}, \& {P{\'e}rez}}]{Isella19}
{Isella}, A., {Benisty}, M., {Teague}, R., {et~al.} 2019, \apjl, 879, L25

\bibitem[{{Larwood} {et~al.}(1996){Larwood}, {Nelson}, {Papaloizou}, \&
  {Terquem}}]{Larwood96}
{Larwood}, J.~D., {Nelson}, R.~P., {Papaloizou}, J.~C.~B., \& {Terquem}, C.
  1996, \mnras, 282, 597

\bibitem[{{Lesur} {et~al.}(2022){Lesur}, {Ercolano}, {Flock}, {Lin}, {Yang},
  {Barranco}, {Benitez-Llambay}, {Goodman}, {Johansen}, {Klahr}, {Laibe},
  {Lyra}, {Marcus}, {Nelson}, {Squire}, {Simon}, {Turner}, {Umurhan}, \&
  {Youdin}}]{Lesur22}
{Lesur}, G., {Ercolano}, B., {Flock}, M., {et~al.} 2022, arXiv e-prints,
  arXiv:2203.09821

\bibitem[{{Lodato} \& {Price}(2010)}]{Lodato2010}
{Lodato}, G., \& {Price}, D.~J. 2010, \mnras, 405, 1212

\bibitem[{{Lubow}(2022)}]{Lubow2022}
{Lubow}, S.~H. 2022, \mnras, 516, 5446

\bibitem[{{Lubow} {et~al.}(2015){Lubow}, {Martin}, \& {Nixon}}]{Lubow15}
{Lubow}, S.~H., {Martin}, R.~G., \& {Nixon}, C. 2015, \apj, 800, 96

\bibitem[{{Lubow} \& {Ogilvie}(2001)}]{Lubow2001}
{Lubow}, S.~H., \& {Ogilvie}, G.~I. 2001, ApJ, 560, 997

\bibitem[{{Martin} \& {Lubow}(2011)}]{Martin2011}
{Martin}, R.~G., \& {Lubow}, S.~H. 2011, \mnras, 413, 1447

\bibitem[{{Miranda} \& {Lai}(2015)}]{Miranda2015}
{Miranda}, R., \& {Lai}, D. 2015, \mnras, 452, 2396

\bibitem[{{Monaghan}(1992)}]{Monaghan1992}
{Monaghan}, J.~J. 1992, \araa, 30, 543

\bibitem[{{Morbidelli} {et~al.}(2014){Morbidelli}, {Szul{\'a}gyi}, {Crida},
  {Lega}, {Bitsch}, {Tanigawa}, \& {Kanagawa}}]{Morbidelli2014}
{Morbidelli}, A., {Szul{\'a}gyi}, J., {Crida}, A., {et~al.} 2014, \icarus, 232,
  266

\bibitem[{{Mosqueira} \& {Estrada}(2003)}]{Mosqueira03}
{Mosqueira}, I., \& {Estrada}, P.~R. 2003, \icarus, 163, 198

\bibitem[{{Nixon} \& {Pringle}(2021)}]{Nixon2021}
{Nixon}, C.~J., \& {Pringle}, J.~E. 2021, \na, 85, 101493

\bibitem[{{Okazaki} {et~al.}(2002){Okazaki}, {Bate}, {Ogilvie}, \&
  {Pringle}}]{Okazaki2002}
{Okazaki}, A.~T., {Bate}, M.~R., {Ogilvie}, G.~I., \& {Pringle}, J.~E. 2002,
  MNRAS, 337, 967

\bibitem[{{Paczynski}(1977)}]{Paczynski1977}
{Paczynski}, B. 1977, ApJ, 216, 822

\bibitem[{{Papaloizou} \& {Pringle}(1977)}]{Papaloizou1977}
{Papaloizou}, J., \& {Pringle}, J.~E. 1977, \mnras, 181, 441

\bibitem[{{Papaloizou} \& {Pringle}(1984)}]{Papaloizou1984}
{Papaloizou}, J.~C.~B., \& {Pringle}, J.~E. 1984, \mnras, 208, 721

\bibitem[{{Pascucci} {et~al.}(2022){Pascucci}, {Cabrit}, {Edwards}, {Gorti},
  {Gressel}, \& {Suzuki}}]{Pascucci22}
{Pascucci}, I., {Cabrit}, S., {Edwards}, S., {et~al.} 2022, arXiv e-prints,
  arXiv:2203.10068

\bibitem[{{Price}(2007)}]{Price2007}
{Price}, D.~J. 2007, \pasa, 24, 159

\bibitem[{{Price}(2012)}]{Price2012a}
---. 2012, Journal of Computational Physics, 231, 759

\bibitem[{{Price} \& {Federrath}(2010)}]{Price2010}
{Price}, D.~J., \& {Federrath}, C. 2010, \mnras, 406, 1659

\bibitem[{{Price} {et~al.}(2018){Price}, {Wurster}, {Tricco}, {Nixon},
  {Toupin}, {Pettitt}, {Chan}, {Mentiplay}, {Laibe}, {Glover}, {Dobbs},
  {Nealon}, {Liptai}, {Worpel}, {Bonnerot}, {Dipierro}, {Ballabio}, {Ragusa},
  {Federrath}, {Iaconi}, {Reichardt}, {Forgan}, {Hutchison}, {Constantino},
  {Ayliffe}, {Hirsh}, \& {Lodato}}]{Price2018}
{Price}, D.~J., {Wurster}, J., {Tricco}, T.~S., {et~al.} 2018, \pasa, 35, e031

\bibitem[{{Pringle}(1991)}]{Pringle1991}
{Pringle}, J.~E. 1991, MNRAS, 248, 754

\bibitem[{{Quillen} \& {Trilling}(1998)}]{Quillen1998}
{Quillen}, A.~C., \& {Trilling}, D.~E. 1998, \apj, 508, 707

\bibitem[{{Schulik} {et~al.}(2020){Schulik}, {Johansen}, {Bitsch}, {Lega}, \&
  {Lambrechts}}]{Schulik20}
{Schulik}, M., {Johansen}, A., {Bitsch}, B., {Lega}, E., \& {Lambrechts}, M.
  2020, \aap, 642, A187

\bibitem[{{Shakura} \& {Sunyaev}(1973)}]{SS1973}
{Shakura}, N.~I., \& {Sunyaev}, R.~A. 1973, A\&A, 24, 337

\bibitem[{{Smak}(1984)}]{Smak84}
{Smak}, J. 1984, \actaa, 34, 93

\bibitem[{{Suffak} {et~al.}(2022){Suffak}, {Jones}, \& {Carciofi}}]{Suffak2022}
{Suffak}, M., {Jones}, C.~E., \& {Carciofi}, A.~C. 2022, \mnras, 509, 931

\bibitem[{{Szul{\'a}gyi} {et~al.}(2016){Szul{\'a}gyi}, {Masset}, {Lega},
  {Crida}, {Morbidelli}, \& {Guillot}}]{Szulagyi2016}
{Szul{\'a}gyi}, J., {Masset}, F., {Lega}, E., {et~al.} 2016, \mnras, 460, 2853

\bibitem[{{Szul{\'a}gyi} {et~al.}(2014){Szul{\'a}gyi}, {Morbidelli}, {Crida},
  \& {Masset}}]{Szulagyi14}
{Szul{\'a}gyi}, J., {Morbidelli}, A., {Crida}, A., \& {Masset}, F. 2014, \apj,
  782, 65

\bibitem[{{Tabone} {et~al.}(2022){Tabone}, {Rosotti}, {Cridland}, {Armitage},
  \& {Lodato}}]{Tabone22}
{Tabone}, B., {Rosotti}, G.~P., {Cridland}, A.~J., {Armitage}, P.~J., \&
  {Lodato}, G. 2022, \mnras, 512, 2290

\bibitem[{{Tanigawa} {et~al.}(2012){Tanigawa}, {Ohtsuki}, \&
  {Machida}}]{Tanigawaetal2012}
{Tanigawa}, T., {Ohtsuki}, K., \& {Machida}, M.~N. 2012, ApJ, 747, 47

\bibitem[{{Ward} \& {Canup}(2010)}]{Ward10}
{Ward}, W.~R., \& {Canup}, R.~M. 2010, \aj, 140, 1168

\bibitem[{{Weidenschilling}(1977)}]{Weidenschilling77aero}
{Weidenschilling}, S.~J. 1977, \mnras, 180, 57

\bibitem[{{Whipple}(1972)}]{Whipple72}
{Whipple}, F.~L. 1972, in From Plasma to Planet, ed. A.~{Elvius}, 211

\bibitem[{{Xu} \& {Goodman}(2018)}]{Xu2018}
{Xu}, W., \& {Goodman}, J. 2018, \mnras, 480, 4327

\bibitem[{{Zhu} {et~al.}(2016){Zhu}, {Ju}, \& {Stone}}]{Zhu2016}
{Zhu}, Z., {Ju}, W., \& {Stone}, J.~M. 2016, \apj, 832, 193

\end{thebibliography}

%% This command is needed to show the entire author+affilation list when
%% the collaboration and author truncation commands are used.  It has to
%% go at the end of the manuscript.
%\allauthors

%% Include this line if you are using the \added, \replaced, \deleted
%% commands to see a summary list of all changes at the end of the article.
%\listofchanges

\end{document}